\documentclass[twocolumn]{aastex631}
\usepackage{multirow}
\usepackage{xcolor}

\begin{document}
\title{Episodic Protostellar Outflows: A Proper Motion Study in Serpens South}

\author[0000-0002-6747-4505]{Simran Kaur}
\affiliation{Department of Astronomy, University of Virginia, Charlottesville, VA 22904, USA}
\affiliation{National Radio Astronomy Observatory, Charlottesville, VA 22903, USA }

\author{Adele L. Plunkett}
\affiliation{National Radio Astronomy Observatory, Charlottesville, VA 22903, USA }

\author[0000-0002-7402-6487]{Zhi-Yun Li}
\affiliation{Department of Astronomy, University of Virginia, Charlottesville, VA 22904, USA}

\author[0000-0002-7993-4118]{Per Bjerkeli}
\affiliation{Chalmers University of Technology, Department of Space, Earth and Environment, 412 96 Gothenburg, Sweden}

\author[0000-0001-5811-0454]{Manuel Fern\'andez-L\'opez}
\affiliation{Instituto Argentino de Radioastronom\'ia (CCT-La Plata, CONICET; UNLP; CICPBA), C.C. No. 5, 1894, Villa Elisa, Buenos Aires, Argentina}
\affiliation{Facultad de Ciencias Astron\'omicas y Geof\'isicas, Universidad Nacional de La Plata, Paseo del Bosque S/N, B1900FWA La Plata, Argentina}

\author[0000-0002-3835-3990]{Jon P.\ Ramsey}
\affiliation{Department of Astronomy, University of Virginia, Charlottesville, VA 22904, USA}
\affiliation{Bluedrop Training \& Simulation, Inc., 36 Solutions Drive \#300, Halifax, Nova Scotia B3S 1N2, Canada}

\author{H\'ector G. Arce}
\affiliation{Department of Astronomy, Yale University, New Haven, CT 06511, USA}

\author[0000-0003-0551-4898]{Mihika Rao}
\affiliation{Department of Astronomy, University of Virginia, Charlottesville, VA 22904, USA}


\begin{abstract}
 Serpens South is a notable protostellar cluster with signatures of recent star formation, including a high protostellar fraction. Within the region, several protostars provide important mechanical feedback via outflows. We investigate the molecular outflow driven by the Class 0 protostar known as CARMA-7, which is located in the central hub of Serpens South. Two epochs of observations with the Atacama Large Millimeter/sub-millimeter Array (ALMA) with sub-arcsecond resolution capture the well-collimated, episodic outflow with distinct ejecta knots. This outflow, traced with $^{12}$CO {\textit{J} = 3 - 2} and SiO {\textit{J} = 8 - 7}, appears to be mostly blueshifted to the south and redshifted to the north, with radial velocities up to at least 25 km s$^{-1}$. We present the first proper motion study of this outflow by comparing observations taken in 2017 and 2021. The measured proper motions trace the ejection history and refine the inferred outflow morphology. We find that proper motions of the $^{12}$CO {\textit{J} = 3 - 2}  and SiO {\textit{J} = 8 - 7} knots range between 10 - 40 $\mathrm{mas}\: yr^{-1}$, corresponding to $30-90$\ km s$^{-1}$.  Using radial velocities and proper motions, we determined that the inclinations of the outflow lobes are $\sim 60^\circ -80^\circ$ relative to the line of sight. Finally, we measure ejection timescales as recent as $\sim$10 to $\sim$120 years, and we discuss this in the context of other known protostars with recent ejecta events.


\end{abstract}



\section{Introduction} \label{sec:intro}
In the process of star formation, a Class 0 protostar accumulates mass by drawing in material from its environment. The growth of protostars indicates that this accumulation of material occurs in episodes rather than continuously. Outbursts can last for several hundred years, while the intervals between outbursts can span thousands of years \citep{2012Stamatellos}.  Studying such phenomena requires finding sites where protostars may have experienced such episodes within recent millennia, or less. 

Previous studies show Serpens South is an infrared dark cloud where very young precluster and protocluster clumps exist \citep{2013Tanaka} located at 440.7 $\pm$ 4.6 pc away from the Earth \citep{Ortizleon2023}. Star formation in Serpens South is embedded in a filamentary dark cloud \citep{2008Gutermuth}. The network of filaments connects at a central hub, which has indications of gas accretion \citep{2014Nakamura,2014Fernandez}. A possible collision between filaments triggered the formation of a central protocluster containing about 50 prestellar and protostellar objects \citep{2014Nakamura,2018Plunkett}.
\cite{2015BPlunkett} showed that CARMA6/7 are two protostars in close proximity within the central region of Serpens South, where their extended C$^{18}$O emission overlap. However, CARMA-7 (hereafter, C7) has brighter emission than CARMA-6 (hereafter, C6) therefore, suggesting different characteristics for the two different sources. 

C7 is a Class 0 protostar, which is consistent with the morphology and characteristics of the collimated outflow \citep{2015BPlunkett}. In molecular outflows, carbon monoxide is the most common molecule whose low-level rotational states can easily be excited \citep{2014Nakamura}. The $^{12}$CO {\textit{J} = 2 - 1} emission of C7 extends from north to south spanning $>80\arcsec$ with a position angle of $4^\circ$ east of north \citep{2015BPlunkett}. Moreover, the clumpy emission from C7 outflows show 22 bright distinct knots of molecular gas that are redshifted and blueshifted to the north and south, respectively, compared with the systemic cloud velocity of 8 km s$^{-1}$\citep{2015APlunkett,2013HelenK}.  Within clustered regions especially, SiO can be a valuable complementary jet tracer to CO for several reasons: it's less likely than CO to be contaminated by envelope and/or outflow emission from neighboring sources; its enhanced abundance and collimated structure can delineate the outflow axis; and its shock-excited nature provokes a well-defined clump-like structure \citep[e.g.~][]{Arce2007, 2017Lee, 2022Lee}.  The CALYPSO IRAM-PdBI survey revealed C7 to be one of the 67\% of the Class 0 protostars in their sample that drives an SiO jet, and several knots are seen with angular resolution of $\sim1^{\prime\prime}$ \citep{Podio2021}.

To understand the kinematics of the outflow from C7, we present a proper motion analysis of the C7 outflow lobes observed with sub-arcsec resolution. To our knowledge, this research is the first effort to acquire the proper motion measurements of C7 outflows, which was done for the period from 2017 to 2021 by observing the two molecules \(^{12}\mathrm{CO}\) {\textit{J} = 3 - 2} (hereafter  $^{12}$CO) and \(\mathrm{SiO}\) {\textit{J} = 8 - 7} (hereafter SiO). Additionally, we determine the inclination angle and dynamical time for each knot to accurately trace the gas kinematics. The methodology of the observations is outlined in \S \ref{sec:Observations and Data}. Results are detailed in \S \ref{sec:Results and Discussion}, where we describe the source reference position, knot proper motions, inclination angles and dynamical times. Discussion of the outflow structure and kinematics of molecular gas within the field is given in \S \ref{sec:discussion}. The main conclusions of our study are summarized in \S \ref{sec:conclusion}.


\section{Observations and Data} \label{sec:Observations and Data}

Two sets of C7 observations were performed with the 12-meter array of the Atacama Large Millimeter/sub-millimeter Array (ALMA) between 2016 August 17 and 2017 May 20 (mean date of 2017 January 2), and again on 2021 May 17. The data were obtained from the ALMA Science Archive (Project IDs: 2015.1.00283.S and 2019.1.00912.S). The 2017 observations were performed in the C-5 configuration, covering a $uv$ range of 15.1m - 612.4m, while the 2021 observations used the C-6 configuration with a $uv$ range of 15.1m - 2500m.\

The 2017 and 2021 observations were made in Band 7. The spectral setup consisted of four basebands with spectral windows centered at 334.975, 336.798, 345.786, and 347.155 GHz. The bandwidths of these basebands were 1873, 234, 469, and 234 MHz, with frequency resolutions of 0.97, 0.24, 0.48, 0.24 MHz, respectively.  The primary focus of this work is the $^{12}$CO $J = 3 - 2$\ transition with a rest frequency of 345.795 GHz, and the SiO $J = 8 - 7$\ transition with a rest frequency of 347.331 GHz. The mean frequency of the line-free channels used for the continuum image is 341.178 GHz, which corresponds to a wavelength of approximately $0.8\,\mathrm{mm}$. Standard catalog calibrators were selected for both the 2017 and 2021 observations. In 2017, J1751+0939, J1830+0619, and Pallas were used as the bandpass, phase, and flux calibrators, respectively. In 2021, J1824+0119 and J1924-2914 served as bandpass calibrators, J1851+0035 as the phase calibrator, and J1924-2914 as the flux calibrator.

We obtained the 2017 and 2021 data cubes from the ALMA Science Archive (ASA). The 2017 and 2021 observations were processed using Common Astronomy Software Applications \citep[CASA,][]{2022CASA} versions 5.6.1-8 and 6.1.1.15, respectively. The molecular line data cubes and continuum images were produced following the standard ALMA pipeline using the CASA \texttt{tclean} task with Briggs weighting and a robust parameter of 0.5. The original angular resolutions of the 2017 (epoch 1) and 2021 (epoch 2) observations were approximately 0.14\arcsec\ and 0.16\arcsec, respectively, for both the molecular line data cubes and continuum images. To compare the molecular line emission between the two epochs, the data cubes were convolved to a common angular resolution of 0.2\arcsec\ (equivalent to 88 au at the distance of the source). The continuum images were smoothed to a common angular resolution of 0.16\arcsec\ ($\sim$ 69 au) and were used to determine the reference position of the C7 protostar. We used a common pixel size of 0.025\arcsec\ (11 au) for all images. We assume an astrometric accuracy of 5--10\% of the synthesized beam, based on the Cycle 3 and Cycle 7 ALMA Technical Handbooks \citep{2015Remijan, 2019Remijan}, which corresponds to 8--16 mas and is smaller than the adopted pixel size. All images were centered at a common phase center of RA = 18:30:04.126, Dec = -2:03:02.72.


We used the sigma-clipped statistics method in the python package astropy to calculate the root-mean-square (rms) noise levels for the $^{12}$CO and SiO data cubes and continuum images. We applied a 3$\sigma$ threshold and excluded data points deviating by more than 3$\sigma$ from the mean as outliers. The rms noise was then calculated from the remaining data; for data cubes, the rms noise was determined per 0.5 km s$^{-1}$\ channel.
For the $^{12}$CO data cubes, the rms noise per channel is 3.4 and 8.3 mJy beam$^{-1}$ for epochs 1 and 2, respectively. Similarly, for the SiO data cubes, the rms noise per channel is 3.2 and 4.5  mJy beam$^{-1}$ for epochs 1 and 2, respectively. The rms noise levels of the continuum images are 1.5 and 1.9 mJy beam$^{-1}$ for epochs 1 and 2, respectively.


\section{Results} \label{sec:Results and Discussion}
\subsection{Reference position} \label{sec:continuum}

\begin{figure*}[ht!]
\centering
    \includegraphics[width=0.9\textwidth, height=0.6\textheight, clip, trim={0 0cm 0 0cm}]
    {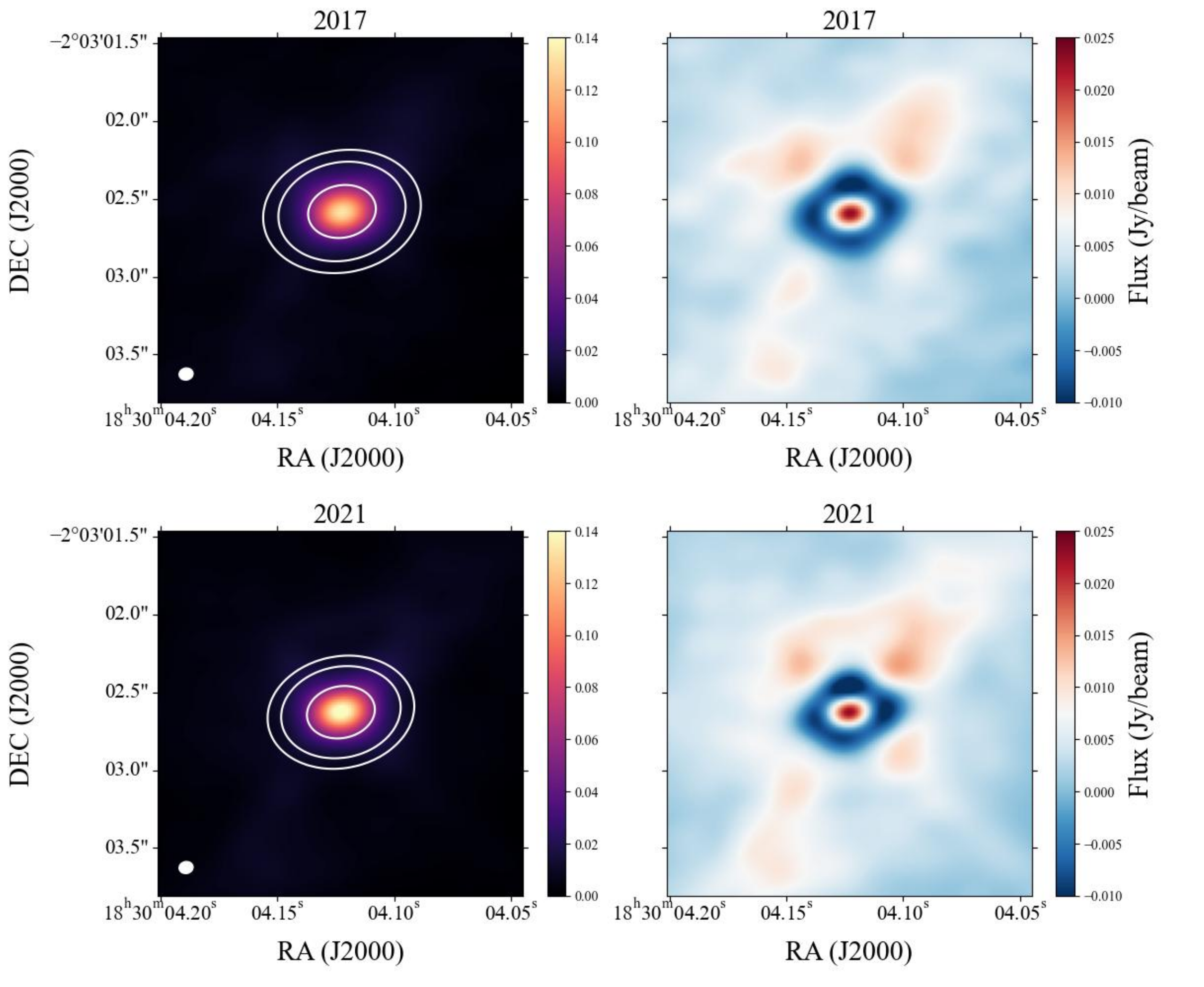}
   
   \caption{\textbf{Top left:} $0.8\,\mathrm{mm}$ continuum emission in 2017 (colorscale) overlaid with contours (white) to represent the 2D Gaussian fit to the data. Contours show levels of 3, 10 and 50$\sigma$, where $\sigma=$ 1.5 mJy beam$^{-1}$ is the rms noise level of the 2017 continuum image. {\textbf{Top right:} Residual map after subtracting the Gaussian model from the 2017 continuum emission. \textbf{Bottom left:} Same as the top left panel, but for the 2021 observations,} with an rms noise level of $\sigma=$ 1.9 mJy beam$^{-1}$.} Gaussian fits were performed on images convolved to a common beamsize of $0.16\arcsec$.
\label{fig:2dgauss}
\end{figure*}

In order to measure proper motions of emission features, we needed to identify a reference position in each image. We fit a 2D Gaussian to the continuum emission in each image using the \texttt{astropy.modeling} \citep{2018Astropy} function \texttt{LevMarLSQFitter}, which incorporates the Levenberg-Marquardt algorithm and least squares statistic package to fit a model to the data. The resulting fits and residual maps are shown in Figure \ref{fig:2dgauss}. We determined that the best-fit Gaussian is centered at RA=18:30:04.122,  DEC=-02:03:02.58 for epoch 1 and RA=18:30:04.123, DEC=-02:03:02.62 for epoch 2, respectively. The difference in positions of the 2D Gaussian fits between the two epochs is 0.043\arcsec almost directly south. The central position of the C7 protostar is consistent with the source labeled as serps-45 by \citet{2018Plunkett}, who reported a location of RA=18:30:04.116, DEC= -2:03:02.51 (with a beam size of 1\arcsec).

To check for a systematic offset between the images, we measured the locations of two continuum sources in the southwest of the field of view in both epochs. These sources correspond to C6 (also referred to as serps33) and serps38 from \cite{2015APlunkett,2018Plunkett}. Both C6 and Serps 38 are located within the diffuse emission region of Serpens South and have slight offsets of 0.04\arcsec\ towards the south between epochs 1 and 2 see figure \ref{fig:12co_channel}. Since this is consistent with the offset measured for C7, we adopt an offset between the two epochs of 43 mas south to represent the cloud proper motion and any astrometric uncertainty in the observations, and we adjust the measured outflow motion by this amount in Sec. \ref{subsec:Gas Dynamics}. 

We also note that \cite{Ortizleon2023} determined the proper motion of the Serpens South cloud to be 0.37 $\mathrm{mas\ yr}^{-1}$ and -6.06 $\mathrm{mas\ yr}^{-1}$ in RA and Dec, based on \textit{Gaia} sources in the Serpens South and nearby W40 regions.  This would imply offsets of 1.6 and -26 mas in RA and Dec over a temporal baseline of 4.3 years between our two epochs.  The offset in RA is negligible, while the sum of the offset in Dec and the estimated astrometric accuracy (8-16 mas) is equivalent to the offset of $\sim40$\ mas south that we measured based on the three continuum sources in our field. 


\begin{table*}
\centering
\caption{Properties of $^{12}$CO {\textit{J} = 3 - 2} Knots}
\label{tab:knots_co}
\begin{tabular}{rcccccccccc}
\hline
\hline
Knot &&\multicolumn{2}{c}{2017}&\multicolumn{2}{c}{Offset (2021-2017)} & Offset & Offset & $\mu$ & i &$v_{total}$\\
&& R.A. & Decl & $\Delta$ R.A. & $\Delta$ Decl\\
&&(J2000)&(J2000) &$(\arcsec)$&$(\arcsec)$&$(\arcsec)$&(au) & ($\mathrm{mas}\: yr^{-1}$) & ($^{\circ}$) & (kms$^{-1}$)\\
\hline
NK1 && 18:30:04.127  & -02:03:02.45 &0& 0.04 & $0.04 \pm 0.025$& $18 \pm 11$ & $9 \pm 6$ & $50 \pm 18$& $25 \pm 9$  \\
NK2 && 18:30:04.129 &-02:03:02.10 &0.05& 0.148& $0.157 \pm 0.025$& $69 \pm 11$ & $36 \pm 6$ & $78 \pm 2$ & $77 \pm 12$\\
NK3 && 18:30:04.133  &-02:03:01.24 & 0.027  &0.148 &  $0.150 \pm 0.025$& $66 \pm 11$ &$35 \pm 6$& $77 \pm 2$ & $74 \pm 12$\\
NK4 && 18:30:04.142 &-02:03:00.24 &0.027 &0.175 & $0.177  \pm 0.025$ & $78 \pm 11$ & $40 \pm 6$ & $79 \pm 2$ & $86 \pm 12$\\
NK5 &&18:30:04.136   &-02:02:59.43 &0& 0.121 & $0.121  \pm 0.025$ & $53 \pm 11$ & $28 \pm 6$ & $75 \pm 3$ & $60 \pm 11$\\
\hline
SK1 &&18:30:04.120  &-02:03:03.26& -0.027 &-0.122&  $0.124 \pm 0.025$ & $55 \pm 11$ & $29 \pm 6$ & $76 \pm 3$ & $62 \pm 12$ \\
SK2 &&18:30:04.113  &-02:03:04.10  &-0.026  &-0.095 & $0.098 \pm 0.025$  &$43 \pm 11$ & $23 \pm 6$ & $72 \pm 4$ & $50 \pm 11$\\ 
\hline
\end{tabular}
\end{table*}

\begin{table*}
\centering
\caption{Properties of SiO {\textit{J} = 8 - 7} Knots}
\label{tab:knots_sio}
\begin{tabular}{rcccccccccc}
\hline
\hline
Knot &&\multicolumn{2}{c}{2017}&\multicolumn{2}{c}{Offset (2021-2017)} & Offset & Offset & $\mu$ & i &$v_{total}$\\
&& R.A. & Decl & $\Delta$ R.A. & $\Delta$ Decl\\
&&(hms)&(dms) &$(\arcsec)$&$(\arcsec)$&$(\arcsec)$&(au) & ($\mathrm{mas}\: yr^{-1}$) &($^{\circ}$) & (kms$^{-1}$)\\
\hline
NK1 &&18:30:04.126   &-02:03:02.49  & 0& 0.04 & $0.04 \pm 0.025$ &$17 \pm 11$& $13 \pm 6$ & $54 \pm 18$&$27 \pm 9$\\
NK2 &&18:30:04.127&-02:03:02.02 &0.026& 0.117 & $0.120 \pm 0.025$ & $53 \pm 11$ & $29 \pm 6$ &$74 \pm 3$ &$60 \pm 11$\\
NK3 &&18:30:04.136  &-02:03:01.29 &0& 0.144 & $0.144 \pm 0.025$ &$63 \pm 11$ &$34 \pm 6$ &$77 \pm 2$ &$71 \pm 12$ \\
NK4 &&18:30:04.145  &-02:03:00.09 & 0 & 0.169& $0.169 \pm 0.025$ &$74 \pm 11$ &$40 \pm 6$ &$79 \pm 2$ & $83 \pm 12$ \\
NK5 &&18:30:04.145  &-02:02:59.42 & 0.026 &0.169 &$0.171  \pm 0.025$ &$75 \pm 11$ &$40 \pm 6$ & $79 \pm 2$ &$84 \pm 12$ \\
\hline
SK1 &&18:30:04.119  &-02:03:03.14 &0& -0.142 & $0.142 \pm 0.025$ & $62 \pm 11$ & $34 \pm 6$ & $74 \pm 3$ & $71 \pm 11$ \\
\hline
\end{tabular}
\end{table*}

\subsection{Proper Motion}
\label{subsec:knots proper motion}
\begin{figure*}[ht!]
\centering
    \includegraphics[width=1\textwidth, height=0.7\textheight, clip, trim={0 4cm 0 0cm}]
    {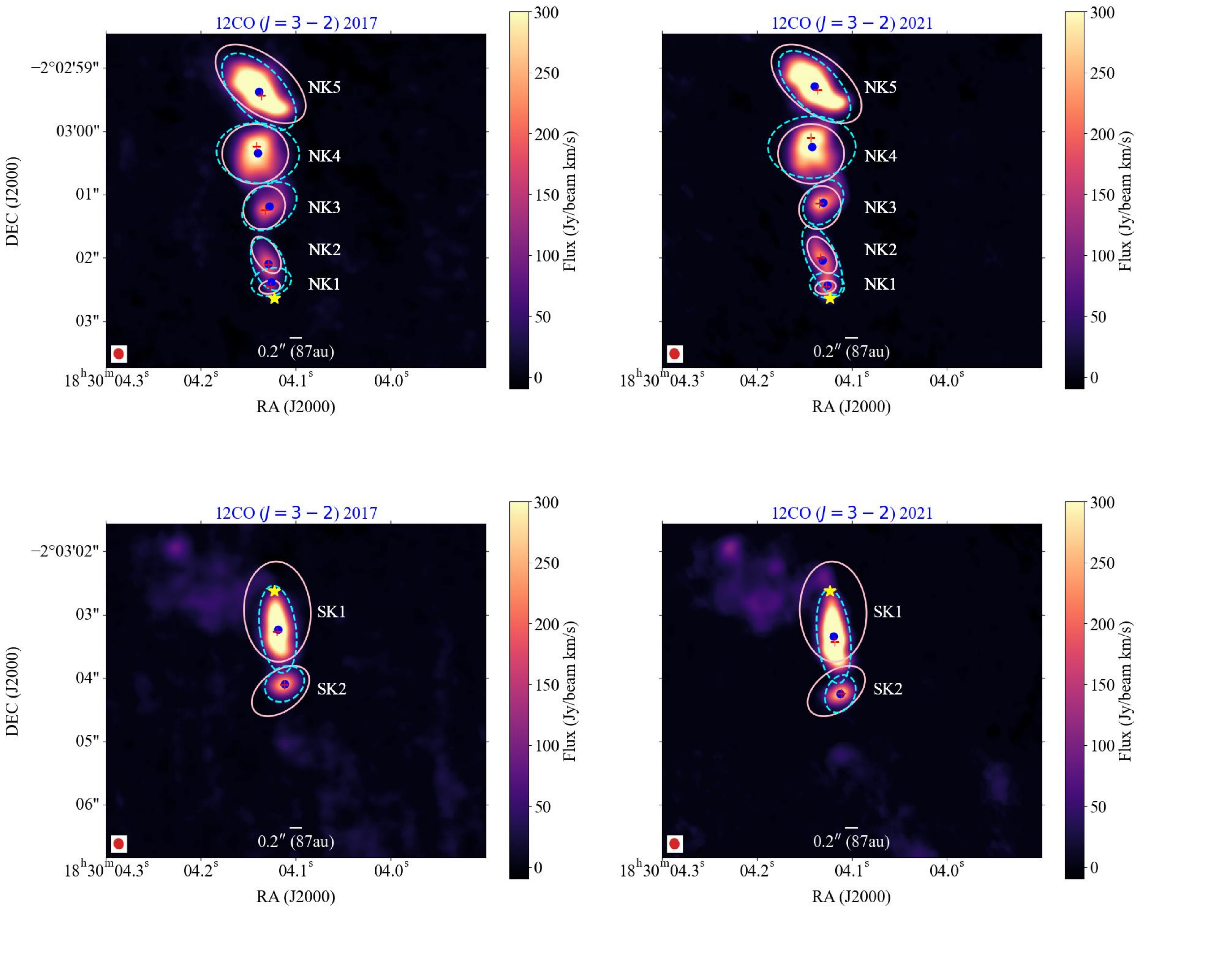}
   \caption{\textbf{Morphology of $^{12}$CO {\textit{J} = 3 - 2} knots in different epochs.} \textbf{Upper row:} Integrated $^{12}$CO intensity moment 0 in 2017 and 2021 (colorscale) presented redshifted knots (ellipse) integrated over velocity channels from 23.5 to 25.5 kms$^{-1}$. \textbf{Lower row:} Same as the upper row but for the blueshifted knots from -9.5 to -7.5 kms$^{-1}$. The location of the protostar (yellow star) and maximum brightness of knots (red $"+"$) are presented in each panel. The centers derived from the 2D Gaussian fits are indicated by blue circles, and the cyan dashed ellipses represent the corresponding Gaussian fits to the knot emission. The beamsize of $\theta=0.2^{\prime\prime}$ is shown with red ellipses in lower left corners of each panel. }
\label{fig:12co_moment0}
\end{figure*}

For each molecular line data cube, we used the Cube Analysis and Rendering Tool for Astronomy \citep[CARTA][]{Comrie2024} to inspect channels individually and determine the velocity ranges where well-defined and spatially coherent emission features above 5$\sigma$\ rms can be seen.  For $^{12}$CO, we made maps of the integrated emission (moment 0) in blueshifted and redshifted channels in the velocity ($v_{ch}$) ranges of -9.5 km s$^{-1}$\ to -7 km s$^{-1}$\ and 23.5 km s$^{-1}$\ to 25.5 km s$^{-1}$. The moment 0 maps from both epochs are shown in Figure \ref{fig:12co_moment0}. We calculated the absolute radial velocity ($v_{flow}$), which represents the motion of each knot relative to the $v_{sys}$ of the cloud. The absolute radial velocity was calculated as $v_{flow}=v_{ch} - v_{sys}$, where $v_{ch}$ is the observed channel velocity and $v_{sys} = 8$ km s$^{-1}$ is the systemic cloud velocity \citep{2015BPlunkett}. We adopted channel velocities of 24 km s$^{-1}$ and $-7$ km s$^{-1}$ for the redshifted and blueshifted $^{12}$CO emission, respectively, resulting in $v_{flow}$ values of 16 km s$^{-1}$ and -15 km s$^{-1}$.


Whereas $^{12}$CO emission is seen over a wider range of velocities (in channels with $v_{ch}$\ ranging from -20 km s$^{-1}$\ to 54 km s$^{-1}$, see Appendix \ref{appendix:channelmaps}), we also notice that the morphology of the emission is velocity-dependent. In general, lower-velocity channels show a less collimated, more expansive morphology with brighter emission than the higher-velocity channels. We selected blueshifted and redshifted ranges that span 2 km s$^{-1}$ each over an intermediate velocity range, where signal-to-noise ratio (SNR) is sufficient to measure several well-defined knots. We adopt half of this range, 1 km s$^{-1}$, as an estimate of the uncertainty in the representative channel velocity ($\sigma_{v_{\mathrm{ch}}}$).

Elliptical regions were drawn by hand around the identifiable knot features in the blueshifted and redshifted moment 0 maps, respectively.  The same elliptical regions are used for both epochs of data.  Within each elliptical region, we identified the location of the pixel which had the peak emission. This is used as a proxy for the location of the knot. In total, we measure the locations of 5 knots in the north (labeled as NK1-5), and 2 knots in the south (SK1, SK2), which are listed in Table \ref{tab:knots_co} and indicated in Figure \ref{fig:12co_moment0}. The continuum source has a systematic offset of $0.043\arcsec$ between the two epochs, with $\Delta \textrm{RA}_{sys}$ = $-0.001\arcsec$ and $\Delta \textrm{Dec}_{sys}$ = $-0.04\arcsec$. Tables \ref{tab:knots_co} and \ref{tab:knots_sio} show $\Delta \textrm{RA} = (\textrm{RA}_{2021})-(\textrm{RA}_{2017}) - \Delta \textrm{RA}_{sys}$ and $\Delta \textrm{Dec} = (\textrm{Dec}_{2021})-(\textrm{Dec}_{2017}) - \Delta \textrm{Dec}_{sys}$.

Based on the knot positions identified in the Epoch 1 and 2 datasets, we determine the positional difference, $D$, for each knot using the formula $D = \sqrt{\Delta\textrm{RA}^2 + \Delta\textrm{Dec}^2}$, expressing the results in both milliarcseconds and astronomical units (refer to Table \ref{tab:knots_co}). The minimum displacement that we could detect in our observations is limited by the pixel size of 0.025\arcsec\ in RA and Dec. While it could be possible to measure proper motions to an accuracy defined by the equation $\sim0.45$HPBW/SNR \citep{Reid1988}, \footnote{HPBW is the half-power beam width of the observations.} since we selected only knots with very high SNR, the corresponding position uncertainty values that we calculated were smaller than the pixel size by more than an order of magnitude. Therefore, we adopt the pixel size of \(0.025\arcsec\) (corresponding to 11 au) as the error in our position measurements.

With these distances, the proper motion, $\mu$, is measured as $\mu = \frac{D}{\Delta\textrm{t}}$. Properties of CO knots are given in Table \ref{tab:knots_co}. The uncertainty in $\mu$ is propagated from the uncertainty in $D$.

A similar procedure is used to measure proper motions of SiO knots, observed at the same time as the $^{12}$CO. Moment 0 maps for blueshifted and redshifted channels were created over the channel velocity ranges of -13.5 km s$^{-1}$\ to -12 km s$^{-1}$\ and 20 km s$^{-1}$\ to 24.5 km s$^{-1}$, respectively. For SiO, representative channel velocities of -12 km s$^{-1}$ and 24 km s$^{-1}$ were adopted for the blueshifted and redshifted components, respectively. The $v_{flow}$ values for the blueshifted and redshifted SiO knots are $-20$ km s$^{-1}$ and 16 km s$^{-1}$, respectively. To verify the representative velocities adopted from the channel maps, we also extracted spatially averaged $^{12}$CO and SiO over the defined knot regions. For the redshifted knots (NK1--NK5), the velocities determined from the peaks of the extracted line profiles range from 13 to 15 km s$^{-1}$. This range is consistent with the representative velocity of 16 km s$^{-1}$ adopted from the channel maps, with differences of only 1--2 km s$^{-1}$ for all knots. For the blueshifted CO and SiO knots (SK1--SK2), the peak velocities of the extracted line profiles range from approximately $-9$ to $-15$ km s$^{-1}$. These values are consistent with the high-velocity channels adopted in our analysis, which spatially coincide with the identified knot structures in the channel maps. Overall, the agreement between the spectral line profiles and the moment maps supports the adopted $v_{\mathrm{ch}}$ values used to determine $v_{\mathrm{flow}}$ as representative of the knot kinematics for the sake of this study.

In addition, the uncertainty in $v_{flow}$ velocity is calculated as follows:

\begin{equation}
\sqrt{\sigma_{v_{\mathrm{ch}}}^{2}
+\sigma_{v_{\mathrm{sys}}}^{2}}
\end{equation}
Assuming that the uncertainty in $v_{\mathrm{sys}}$ is negligible and adopting $\sigma_{v_{\mathrm{ch}}}=1$ km s$^{-1}$ (see \S 3.2), we obtain $\sigma_{v_{\mathrm{flow}}}=1$ km s$^{-1}$ for both $^{12}$CO and SiO. Using these adopted velocities, our results show that displacements of the $^{12}$CO and SiO knots are similar and ranging from $0.04\arcsec - 0.18\arcsec$, with an average displacement of $\sim0.1\arcsec$. In physical units, this is equivalent to displacements of about 40 au on average, and up to about 80 au.

As a consistency check, we measured knot positions using two-dimensional Gaussian fitting in CARTA (v5.1.0) with the image fitting tool. We applied this method to the same blueshifted and redshifted $^{12}$CO integrated emission maps described above to ensure a direct comparison with the peak-brightness measurements. The Gaussian fitting was performed on the identified knots to obtain centroid positions. A similar analysis was performed for the SiO moment 0 maps. The positional offsets derived from the Gaussian-fit centroids were compared with those obtained from the peak-brightness method. We find that the offsets measured using both techniques are consistent. For $^{12}$CO the differences typically range between 0.05\arcsec - 0.06\arcsec, while for SiO even smaller, ranging between 0\arcsec - 0.02\arcsec. These differences are comparable to or smaller than our estimated minimum measurable displacement of 0.04\arcsec over 4.3 years, indicating that the measured proper motions are robust and are not significantly affected by the adopted position-measurement method. The minimum proper motion measured for the knot is $\sim$ 13 $\mathrm{mas}\:yr^{-1}$, corresponding to a displacement of about 17 au over a 4.3 years baseline. We give these values as upper limits in the case of NK1 in Tables \ref{tab:knots_co} and \ref{tab:knots_sio}.

\begin{figure*}[ht!]
\centering
    \includegraphics[width=1\textwidth, height=0.7\textheight, clip, trim={0 4cm 0 0cm}]
    {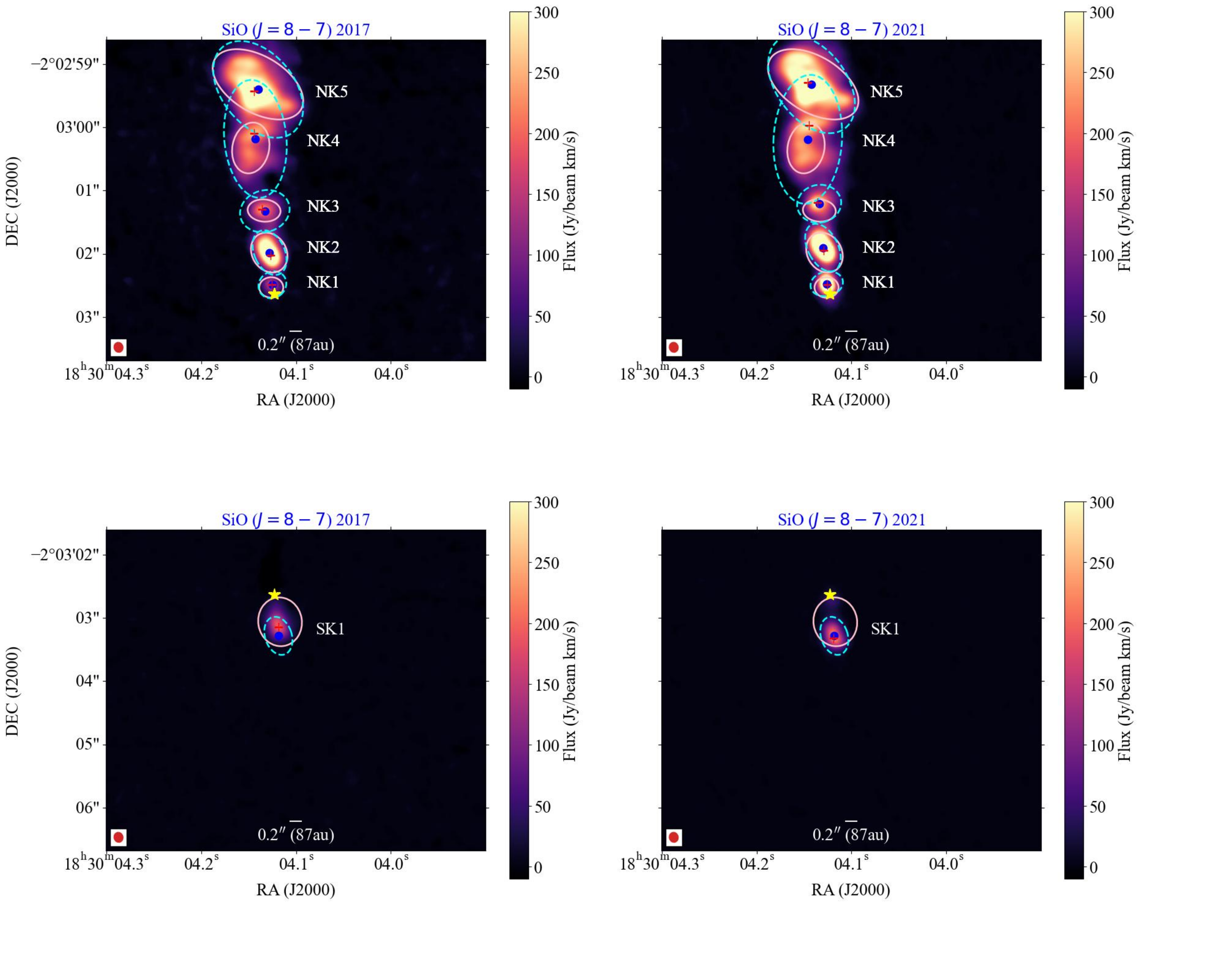}
   \caption{\textbf{Morphology of SiO {\textit{J} = 8 - 7} knots in different epochs.} \textbf{Upper row:} Integrated SiO intensity moment 0 in 2017 and 2021 (colorscale) presented redshifted knots (ellipse) 
 integrated over velocity channels from 20 to 24.5 kms$^{-1}$ \textbf{Lower row:} Same as the upper row but for the blueshifted knots from -13.5 to -12 kms$^{-1}$. The location of the protostar (yellow star) and maximum brightness of knots (red $"+"$) are presented in each panel. The centers derived from the 2D Gaussian fits are indicated by blue circles, and the cyan dashed ellipses represent the corresponding Gaussian fits to the knot emission. The beamsize of $\theta=0.2^{\prime\prime}$ is shown with red ellipses in lower left corners of each panel.}
\label{fig:sio_moment0}
\end{figure*}

\begin{figure*}[ht!]
\centering
    \includegraphics[width=1\textwidth, height=0.6\textheight, clip, trim={0 4cm 0 0cm}]
    {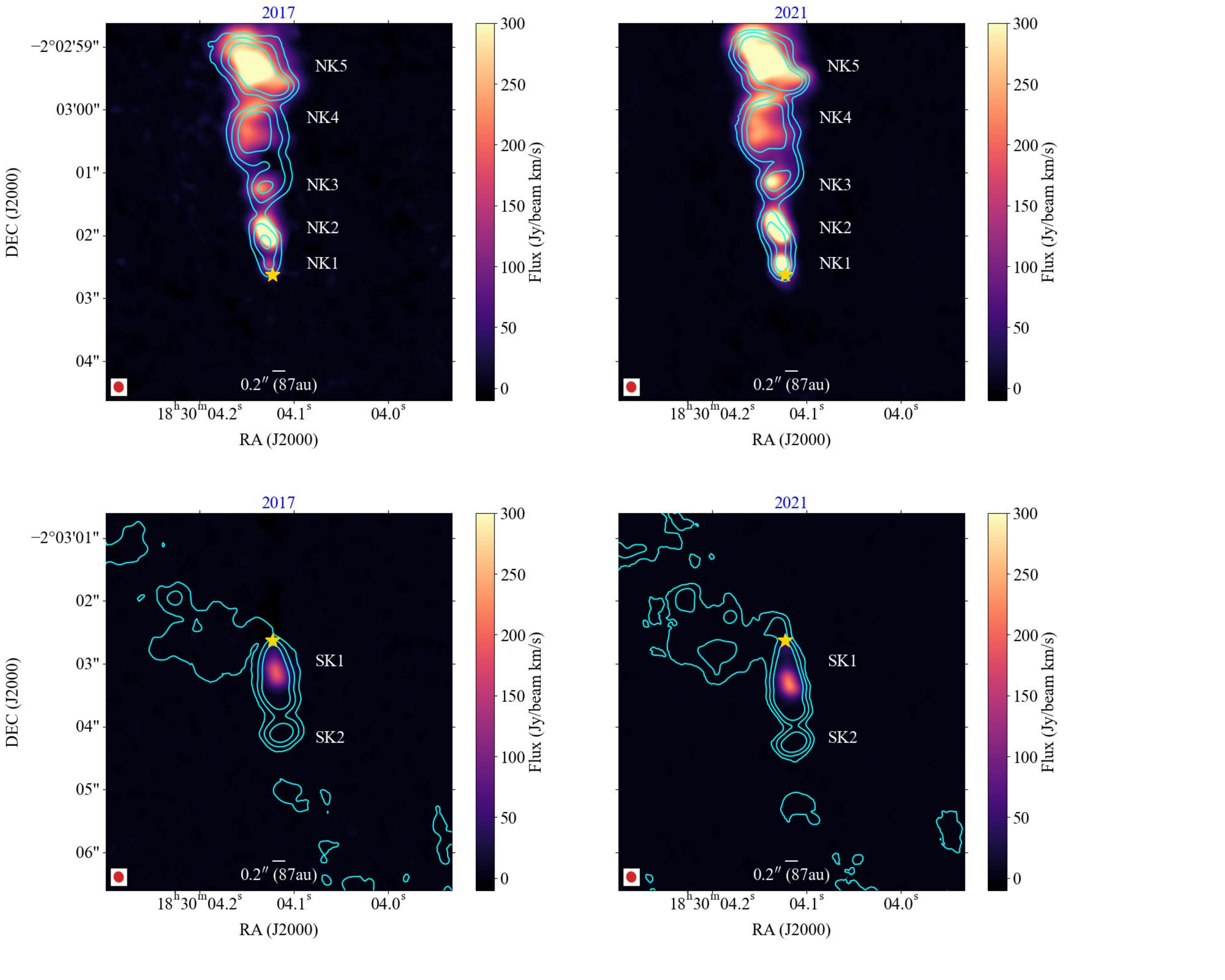}
   \caption {\textbf{Spatial comparison of the $^{12}$CO and SiO emission toward the C7 protostellar outflow.} \textbf{Upper row:} SiO integrated intensity (moment 0) maps of the redshifted emission from the 2017 and 2021 observations, overlaid with $^{12}$CO contours (cyan) begin with 7$\sigma$, 21$\sigma$, and 50$\sigma$. \textbf{Lower row:} same as above, but for the blueshifted emission. The location of the protostar (yellow star) is presented in each panel. The beamsize of $\theta=0.2^{\prime\prime}$ is shown with red ellipses in lower left corners of each panel.}

\label{fig:ratioplot}
\end{figure*}

\subsection{Inclination angle of the CARMA-7 Knots}
\label{subsec:Inclination}

We used the proper motions to measure the inclination angle with respect to the line of sight, where an $i=90^\circ$ outflow lies on the plane of the sky. The outflow inclination angle can be measured using the following equation: 
\begin{equation}
    \tan (i) = \left| \frac{\mu}{v_{flow}} \right|
\end{equation} where $\mu$ represents the motion of each knot in the plane of the sky (in units of km s$^{-1}$), and $v_{flow}$ is the radial velocity of a knot relative to the protostar. We calculate the error of the inclination angle using the following equation:
\[
\sigma_i =
\frac{
\sqrt{
\left(\frac{\sigma_\mu}{v_{\mathrm{flow}}}\right)^2
+
\left(\frac{\mu\sigma_{v_{\mathrm{flow}}}}{v_{\mathrm{flow}}^2}\right)^2
}
}
{
1+\left(\frac{\mu}{v_{\mathrm{flow}}}\right)^2
}
\]

\noindent where $\sigma_i$ is the uncertainty in the inclination angle, $\sigma_\mu$ is the uncertainty in the proper motion and $\sigma_{v_{\mathrm{flow}}}$ is the uncertainty in the absolute radial velocity.

We measured inclination angles for the $^{12}$CO and SiO knots for both epochs to investigate trends among a given molecule, as well as to compare inclination angles measured at the same location using the two different molecules. Both $^{12}$CO and SiO  knots have inclination angles ranging between $50^\circ - 80^\circ$, as shown in Tables \ref{tab:knots_co} and \ref{tab:knots_sio}. The inclination angles of blueshifted and redshifted knots are symmetric, showing comparable magnitudes with opposite signs. This is expected for a bipolar outflow, where one lobe is tilted toward the line of sight and other lobe is tilted away.

\subsection{Dynamical Time for CARMA-7 Knots}
\label{subsec:Dynamical Time}

We assessed the dynamical time $(t_{dyn})$ for the knots of $^{12}$CO and SiO using the formula: 
\begin{equation} 
t_{dyn} = \frac{d_{knot}}{\mu},
\end{equation}
where d$_{knot}$\ is the distance of each knot relative to the continuum position of 2017 (Epoch 1, see \S 3.1).  

The dynamical times for the $^{12}$CO and SiO knots span from \textbf{$\sim$}10 to \textbf{$\sim$}120 years, as shown in Tables  \ref{table 6:Dynamical Time 12co} and \ref{tab:Dynamical time SIO}. These calculations are based on the assumption that the knots have been traveling at a constant speed in the plane of the sky.  Even if this did not hold -- for example, if knots decelerated with time -- then we could still compare (estimated) dynamical times of $^{12}$CO and SiO knots if we assumed that they were traveling co-spatially.  In some knots, they do appear cospatial which leads us to believe this assumption is at least partially true.  Conversely, if the $^{12}$CO was ejected from the vicinity of the protostar or composed of entrained ambient material, while the SiO was excited in shocks, then the assumptions for measuring dynamical time may differ. 

\begin{table*}
\centering
\caption{Dynamical Time of $^{12}$CO {\textit{J} = 3 - 2}} Knot
\label{table 6:Dynamical Time 12co}

\begin{tabular}{rcccccccc}
\hline
\hline
Knot&& Distance && $t_{dyn}$\\
&& (\arcsec) && ($yr$)\\

\hline
NK1 && 0.14&& \nodata\\
NK2 &&0.49&& 14\\
NK3 &&1.35&& 39\\
NK4 &&2.36&&58\\
NK5 &&3.16&&115\\
\hline
SK1 &&0.68&&24\\
SK2 &&1.52&&67\\
\hline
\end{tabular}
\end{table*}

\begin{table*}
\centering
\caption{Dynamical Time of SiO {\textit{J} = 8 - 7} }
\label{tab:Dynamical time SIO}

\begin{tabular}{rcccccccc}
\hline
\hline
Knot&& Distance && $t_{dyn}$\\
&& (\arcsec) && ($yr$)\\

\hline
NK1 && 0.10&& \nodata\\
NK2 &&0.56&& 21\\
NK3 &&1.30&& 40\\
NK4 &&2.50&& 65\\
NK5 &&3.17&&81\\
\hline
SK1 &&0.55&& 17\\
\hline
\end{tabular}
\end{table*}

\subsection{Nearby protostellar outflows within the field}
Within the field of view, C7 is accompanied by two neighboring sources located towards the southwest: the previously mentioned C6, and a source we refer to as serps-38 \citep[see][]{2018Plunkett}. Whereas continuum emission has been previously detected for these sources \citep{2015APlunkett,2018Plunkett}, to our knowledge this is the first time that an outflow is identified from serps-38, reinforcing its classification as a protostar. CO emission from serps-38 appears as a blueshifted outflow lobe northeast from the source that spans $\sim10-15^{\prime\prime}$.  It is especially apparent in channels with velocities of $-2$ to 5 km s$^{-1}$ (Figure \ref{fig:12co_channel}; these channels are blueshifted with respect to the cloud velocity of 8 km s$^{-1}$), and it overlaps in projection with the main southern outflow lobe of C7. To the south of serps-38, faint emission is seen in low-velocity redshifted channels, but the proximity to the edge of the map degrades the signal-to-noise. 

Observations of CO also reveal a wide-angle bipolar shell-like structure from C6 with the northern lobe mostly redshifted and the southeast lobe mostly blueshifted.  Gas in the northern lobe reaches radial velocities greater than about 30 km s$^{-1}$, and the southern lobe has lower velocities, although it also extends more substantially beyond the field of view. 

For both C6 and serps-38, we detected outflows in CO but not in SiO.  While not the immediate focus of this study, the outflows from these sources provide additional context for the clustered nature of the environment around C7.

\section{Discussion} \label{sec:discussion}
\subsection{Gas Dynamics In CARMA-7}
\label{subsec:Gas Dynamics}
We observe a bipolar outflow from C7 with distinct ejecta knots. The outflow is morphologically similar to previous findings for HH34 \citep{2012Raga} 
although that source was studied using optical (Hubble Space Telescope) observations, while our study is performed in the sub-millimeter wavelength.  Studies of each of these sources suggest non-constant outflow morphology, and possibly episodic ejection mechanisms. Other sources for which well-defined knots have been measured include HH211 and HH212 \ \citep{2021Jhan,2015Lee,2024Lopez} both of which were observed with ALMA. As for HH211 and HH212, we observed ejecta knots for two molecular tracers, SiO and CO, in the case of C7. 
SiO emission is more suitable for tracing high density material and shocks in the jets. In comparison, CO traces both high density jets and low density outflows \citep{2024Dutta}. 

In Figure \ref{fig:ratioplot}, we present SiO emission overlaid with contours of $^{12}$CO emission. SiO and $^{12}$CO generally exhibit similar spatial distributions along the outflow in the velocity ranges inspected here, although their detailed morphologies vary among individual knots. The $^{12}$CO and SiO emission peaks associated with the identified knots do not exhibit a significant spatial displacement from each other. The positions of the $^{12}$CO and SiO knots are consistent within the measurement uncertainties. While the locations of peak $^{12}$CO emission for some knots are located closer to the protostar than those of SiO knots, others show the opposite trend, with the locations of peak SiO emission located closer to the protostar. This mixed behavior of the knots indicates a more complex internal structure of the outflow, where $^{12}$CO and SiO emission may trace different physical components within the same knot. The difference in peak positions are likely driven from the different temperature and density regimes of the knots, although further analysis is needed to confirm this. \cite{2024Dutta} showed that in some sources, $^{12}$CO and SiO can trace different components of protostellar outflows, with SiO emission associated with narrow, collimated jets and $^{12}$CO emission tracing a wider entrained outflow. In the case of C7, we find that both $^{12}$CO and SiO emission trace high-velocity gas associated with collimated knot structures within the outflow. This difference suggests that different protostellar outflows can exhibit distinct $^{12}$CO and SiO emission behaviors, likely reflecting differences in their outflow structures and evolutionary conditions.

To further investigate jet kinematics we calculated the proper motion of each C7 knot and combined with the corresponding radial velocity ($v_{flow}$) to find the three dimensional velocity or total velocity (v$_{total}$). We used the following formula (v$_{total}$ = $\sqrt{{\mu}^2 + v_{flow}^2}$), and results are given in Tables \ref{tab:knots_co} and \ref{tab:knots_sio}. The uncertainties are also given by:
\begin{equation}
\sigma_{ v_{total}}
=
\frac{
\sqrt{
(\sigma_{\mu}\mu)^2
+
(v_{\mathrm{flow}}\sigma_{v_{\mathrm{flow}}})^2
}
}{
\sqrt{
\mu^2 + v_{\mathrm{flow}}^2
}
}
\end{equation}
The 3 dimensional velocities of $^{12}$CO and SiO knots range between 30 - 90 kms$^{-1}$. Our results show CO and SiO knots move at different velocities and no uniform trend is observed between the SiO and CO knots.



As presented by \cite{2024Dutta}, there are several plausible scenarios to describe SiO formation in these jets: one in which SiO originates through grain sputtering that occurs at a shock region along the jet, another in which SiO could be formed from a jet-launch region within the dust sublimation radius of the protostar itself \citep{1997Caselli,2012Panoglou,1997Schilke}. Therefore, we can observe SiO emission to trace the mass loss in C7. In the case of the dust sputtering scenario, velocities of $\gtrsim 50$\ km s$^{-1}$\ are necessary, along with high preshock densities.  One reason why our determination of v$_{total}$\ is important is that it allows us to confidently presume that such jet velocities are indeed present in this system without substantial assumptions about inclination. This confirms that dust sputtering remains a viable mechanism for SiO formation in the C7 jet. Alternatively, fast, collimated, dust-free disk winds destroy silicate dust grains. The sublimation of silicon into the gas phase causes an SiO shock which can be traced through SiO emission. Our observations of the C7 outflows support the presence of “fast, collimated” wind \cite{2020Tabone} but follow-up work would be necessary to determine the wind temperature and density.

Furthermore, \citet{2024Dutta} illustrate how the SiO and CO trace slightly different components of the same knot, with the SiO and CO found in the middle, faster layer, while $^{12}$CO is also found in the forward and backward fronts \citep[see Fig.~11 in][]{2024Dutta}. 
However, our results suggest that CO and SiO originate from the same knot components in at least some velocity ranges.

\subsection{Episodic accretion and ejection}
\label{subsec:Episodic accretion and ejection}

We calculated the dynamical time $(t_{dyn})$ of the C7 knots to understand the outflow launch history. The estimated dynamical time of the different ejections in the north and south C7 outflows range between \textbf{$\sim$}10 - \textbf{$\sim$}120 years (see Tables \ref{table 6:Dynamical Time 12co}-\ref{tab:Dynamical time SIO}), and our results are consistent with timescales corresponding to another Class 0 outflow, IRAS 15398-3359 \ \citep{2021Vazzano,2024Guzm}. The timescales of the C7 outflow were previously presented in \cite{2015BPlunkett}, but without proper motions to determine the displacement in the plane of the sky, nor a robust correction for inclination angle. They determined the timescale of the most recent ejecta to be around 100 years, which corrected for the blueshifted lobe inclination angle of $\sim80^\circ$ agrees with the timescales we present here. After correcting for the motion of the Serpens South cloud we found the dynamical time for CO and SiO is consistent with each other, indicating a symmetrical outflow for C7. These short dynamical timescales provide a unique opportunity to probe the recent mass-loss history of the protostar. Compared to the lifetime of a protostar of about half a million years, the ability to study gas motions on timescales of a few tens of years offers further constraints on how mass accumulates during the protostellar lifetime. During the early stages of star formation the protostar ejects material episodically which links back to the episodic accretion, which is often much more challenging to observe due to being deeply embedded in the envelope. The C7 protostar is relatively unique because its most recent ejection event is less than 20 years ago, which may be triggered by a recent accretion event as well. \cite{2018Vorobyov} motivated the potential link between accretion and ejection by presenting the corresponding timescales between their accretion model -- specifically clustered bursts with timescales between bursts of a few $\times(10–10^2)$\ yr -- and the C7 outflow ejection events. 

The inclination angles of the redshifted $^{12}$CO and SiO knots (NK1–NK5) exhibit a trend similar to that of the blueshifted $^{12}$CO and SiO knots (SK1–SK2). The similarity in both the spatial distribution and kinematics of the blue- and redshifted knots suggests that the bipolar outflow lobes may have been launched during the same episodic accretion events. While we cannot establish a unique one-to-one correspondence between individual northern and southern knots based on the inclination angles alone, the similarity between the inclination angles among several knots suggest a comparable outflow geometry between the two lobes rather than a direct identification of individual knot pairs. The fact that the overall inclination angles of the outflow determined by constraints from the modeled accretion timescales agree with the inclination angles measured by our work further reinforces the accretion-ejection connection.

To our knowledge, only a handful of other sources seem to have a more recent outflow ejection event reported in the literature. The source B335 presented by \cite{2023Evans} has likely experienced an ejection event in the last 5 years, and MMS 1/OMC-3 studied by \cite{2024Takahashi} using ALMA has an outflow with four distinct knots with timescales between $\sim$5-50 years. Properties of knots in HH212 indicate that molecular gas in the jet was launched 15-30 years ago \citep{Podio2015}.   Recently, JWST observations of IRAS 15398-3359 by \cite{2025ApJ...982..149O} revealed an outflow ejection event as little as 10 years ago. It is reasonable to postulate that these sources have experienced recent episodic ejection events, as for C7. As more high-resolution, high-sensitivity observations become available in the millimeter-wavelength to IR-wavelength regimes, our understanding of the scenarios in which distinct ejecta are present will improve. Whereas protostellar evolution takes on the order of $10^5-10^6$ years to occur, ejection events are the observational evidence of the ongoing accretion processes in the star formation story.

\section{Conclusion} 
\label{sec:conclusion}

We studied the evolution of star formation in the dense, filamentary central region of Serpens South. In this star cluster, the young source C7 is in an early stage of formation, and outflows are beginning to emerge.\\

We measured the proper motions of the C7 outflow features using ALMA datasets from the 2017-2021 epochs. C7 outflows appear episodic, and distinct ejecta knots were found in both the north and the south directions. We found the proper motion of the $^{12}$CO and SiO knots ranges between 10 - 40 $\mathrm{mas}\: yr^{-1}$. With proper motions, we measured the inclination angle ranges between $50^\circ - 80^\circ$ and ultimately dynamical time between \textbf{$\sim$}10 - \textbf{$\sim$}120 years of each knot. In addition, the 3 dimensional velocities of $^{12}$CO and SiO knots range between 30 and 90 km s$^{-1}$. The episodic accretion–ejection events traced by such outflow knots around protostars help us understand the formation history of protostars.


\appendix
\section{Channel maps} \label{appendix:channelmaps}
In this section, we present  maps of the $^{12}$CO and SiO, in order for velocity-dependent features to be explored. For the sake of space, in each panel we integrated over several channels spanning 2 km s$^{-1}$, within which mostly coherent kinematic structures are seen. We remind the reader that the actual channel width of the cubes is 0.5 km s$^{-1}$. Proper motions, as described in \ref{subsec:knots proper motion}, were measured based on the moment maps presented in Figs. \ref{fig:12co_moment0} and \ref{fig:sio_moment0}.

\setcounter{figure}{0}
\renewcommand{\thefigure}{A.\arabic{figure}}
\begin{figure*}[ht!]
    \includegraphics[width=1\textwidth]
    {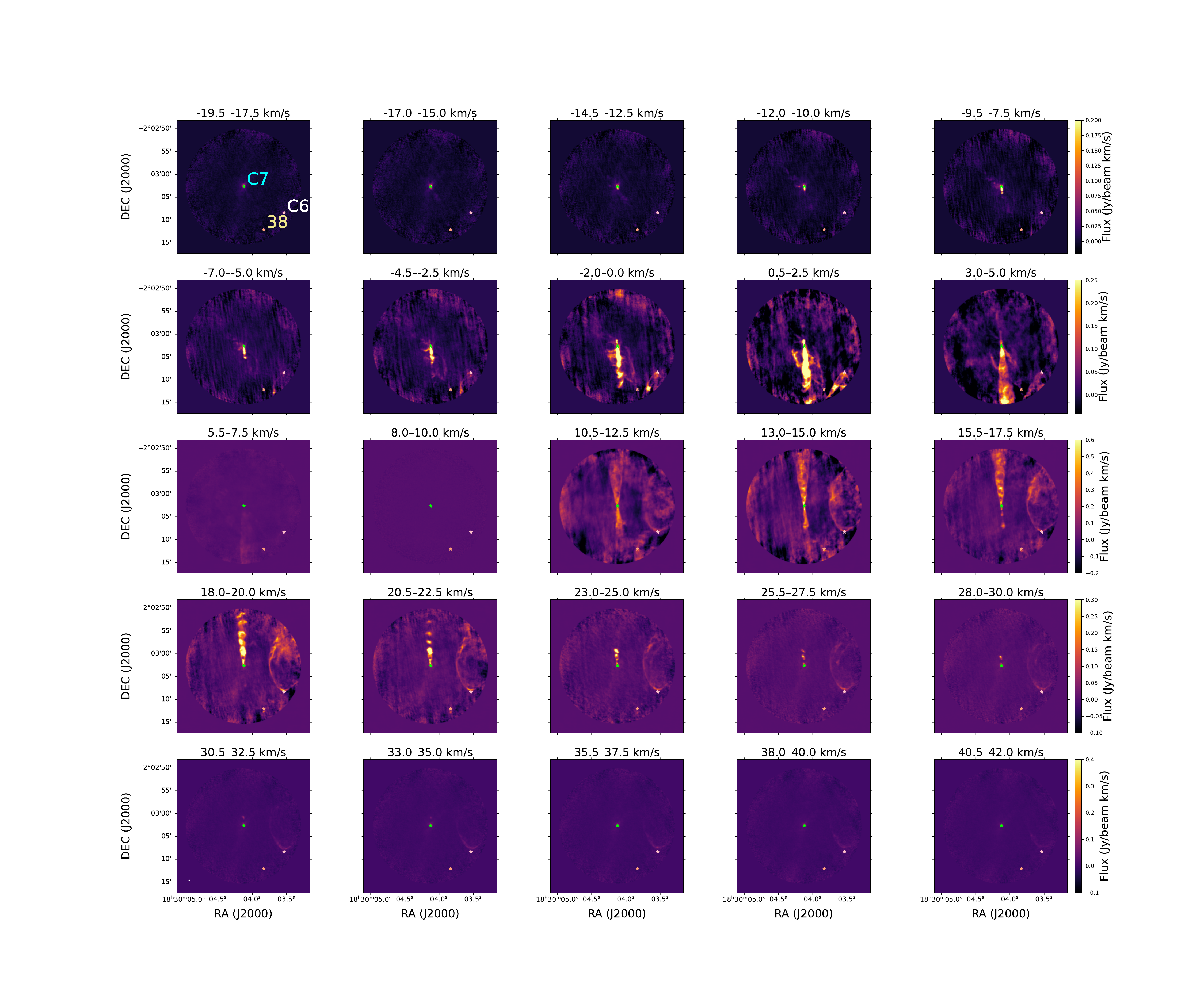}
   \caption{$^{12}$CO maps spanning the velocity range of -19.5 km s$^{-1}$ to 42 km s$^{-1}$ where emission is seen. Each panel shows a moment 0 map made by integrating over five 0.5 km s$^{-1}$ channel maps, and the labels indicate the range of central velocities of channels that were included. The panel that spans 8 km s$^{-1}$\ represents the systemic cloud velocity of the driving source; panels with lower velocities (the upper panels) portray blueshifted gas, and panels with higher velocities (the lower panels) portray redshifted gas, relative to the source. The synthesized beam size (white ellipse) is shown in the lower left panel. The locations of the C7 (lime star), C6 (pink star) and serps38 (orange star, labeled simply “38”) are shown in all panels and labeled  in the first panel for clarity. A common colorbar assigned to the each row, with panels representing different velocity ranges.}
\label{fig:12co_channel}
\end{figure*}

\begin{figure*}[ht!]
    \includegraphics[width=1\textwidth,height=0.7\textheight, clip, trim={0 2cm 0 0cm}]
    {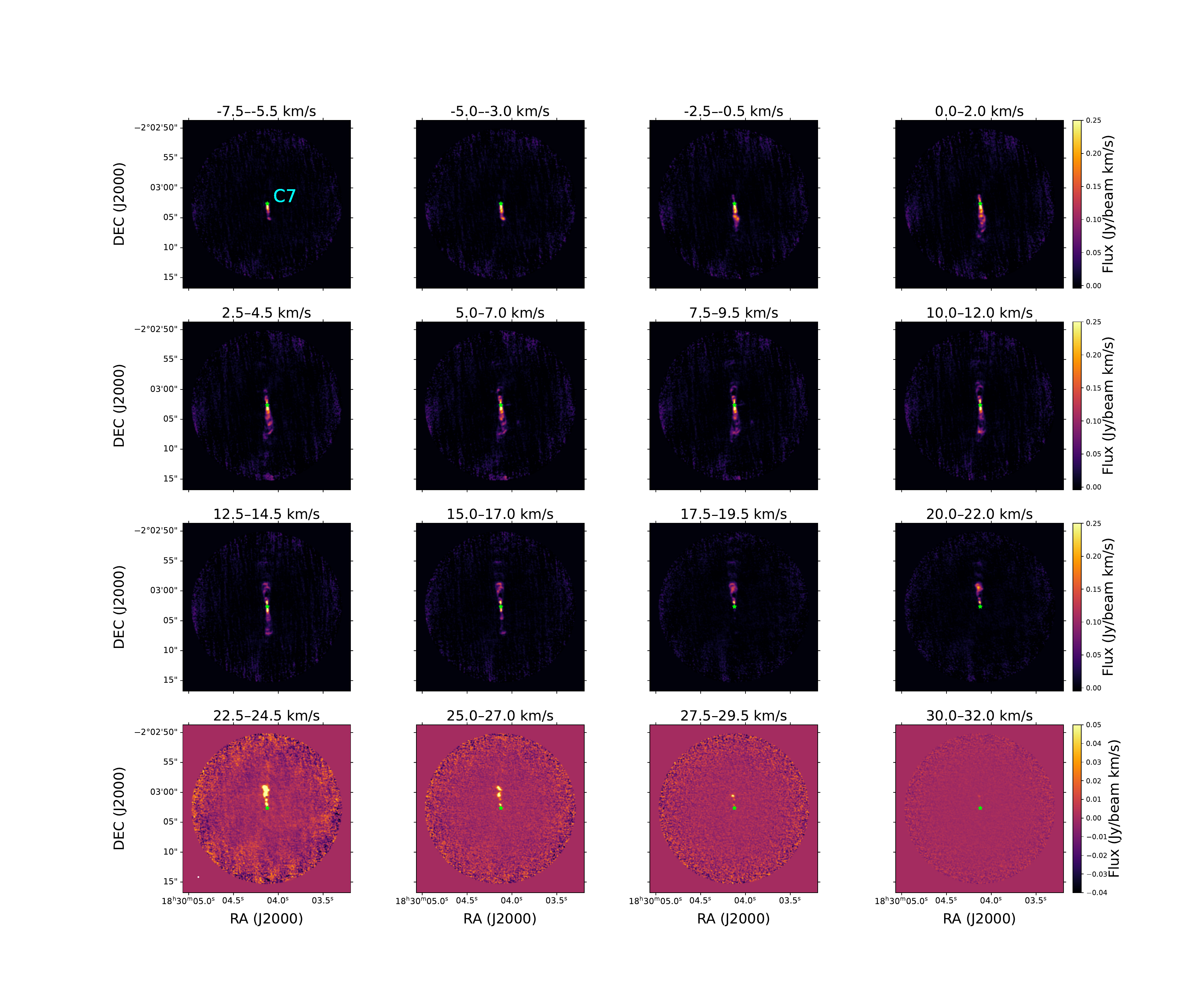}
   \caption{SiO maps spanning the velocity range of -7.5 km s$^{-1}$ to 32 km s$^{-1}$. Each panel shows a moment 0 map made by integrating over five 0.5 km s$^{-1}$ channel maps, and the labels and velocities follow the same convention as in Figure \ref{fig:12co_channel}. The synthesized beam size (white ellipse) is shown in the lower left panel. The location of the C7 protostar (lime star) presented in the first panel. A common colorbar assigned to the each row, with panels representing different velocity ranges.} 
\label{fig:sio_channel}
\end{figure*}


\bibliography{sample631}{}
\bibliographystyle{aasjournal}



\end{document}